# Global View of Bionetwork Dynamics: Adaptive Landscape

Ping Ao

*Department of Mechanical Engineering and Department of Physics*
*University of Washington, Seattle, WA 98195, USA;*
*Shanghai Center for Systems Biomedicine, Shanghai Jiao Tong*
*University, Shanghai, PR China*

February 5 (2009)

**Abstract**

Based on our recent work here I give a nontechnical brief review of a powerful quantitative concept in biology, adaptive landscape. This concept was initially proposed by S Wright 70 years ago, re-introduced by one of the founders of molecular biology and by others in different biological contexts. It was apparently forgotten by mainstream modern biologists for many years. Currently, this concept has found its increasingly important role in the development of systems biology and the modeling of bionetwork dynamics, from phage lambda genetic switch to endogenous network of cancer genesis and progression. It is an ideal quantify to describe the robustness and stability of bionetworks. I will first introduce five landmark proposals in biology on this concept, to demonstrate the important common thread in its theoretical biology development. Then I will discuss a few recent results, focusing on the work showing the logical consistency of adaptive landscape. From the perspective of a working scientist and of what needed for a dynamical theory when confronting empirical data, the adaptive landscape is useful both metaphorically and quantitatively and has captured an essential aspect of biological dynamical processes. Still, many important open problems remain to be solved. Though at the theoretical level the adaptive landscape must exist and it can be used across multiple hierarchical boundaries in biology, many associated issues are indeed vague in their initial formulations and their quantitative realizations are not easy, which are good research topics for quantitative biologists. I will discuss three types of open problems associated with adaptive landscape in a broader perspective.
( Please refer the paper as: Global View of Bionetwork Dynamics: Adaptive Landscape. Ping Ao.
   ***Journal of Genetics and Genomics***. *V.36, 63-73 (2009)*
http://www.jgenetgenomics.org/qikan/epaper/zhaiyao.asp?bsid=14881 )

*"There may still be a use for people who believe there is more in life and in biology than the applied biochemistry of the nucleic acids, always provided that they pay due regard to the man who has been trained to wield modern methods with precision and apply modern logical and mathematical facilities to the interpretation of his results."*

Franck Macfarlane Burnet, 1899-1985

**I. Introduction**

With the emergence of systems biology, the demands on quantitative handling of data become increasingly great (Hood, 2003; Auffray *et al.*, 2009). There have been two general and opposite methodologies available and have been very helpful in facilitating the progress. The statistical analyses which generally focus on data sets themselves, supplemented by biological understandings (Aloy and Russell, 2008; Han, 2008). The mechanistic modeling, on the other hand, focuses on the working of the biological phenomena, at suitable levels of physics, chemistry, and biology, supplemented by statistical analysis (Li *et al.*, 2004; Zhu *et al.*, 2004; Ptashne, 2004; Auffray and Nottale, 2008). The purpose of present overview is, however, on a middle ground approach: A stochastic process approach which can smoothly connect the known two methodologies (Zhu *et al.*, 2007). This middle-ground approach have been applying recently to small bionetworks such as genetic switches (Black *et al.*, 2003; Chabot *et al.*, 2007; Raser and O'Shea, 2005; Zhu *et al.*, 2004) and to those of complex diseases such as cancer (Ao, 2007; Ao *et al.*, 2008) and evolutionary processes (Ao, 2005a; Kussell and Leibler, 2005). Even at metabolic and physiological level, it shows a promising potential in bring out salient biological properties (Ao, 2005b; Ao *et al.* 2008b; Elf *et al.*, 2007; Hanson and Schnell, 2008; Lee *et al.*, 2007; Qian *et al.*, 2003; Scott *et al.*, 2007).

The central concept in such stochastic dynamics approach is the adaptive landscape. It is different from the direct real time calculation or simulation type (Gillespie, 2007) in that it aims to get middle and long time behaviors. It differs from those focusing on moments (Gadgil *et al.,* 2006) but leans toward to those of nonequilibrium thermodynamics approaches (Qian, 2005). It also differs from other more formal stochastic approaches in biology but more from mathematical point of view (Malyshev and Pirogov, 2008). The powerful adaptive landscape concept was first proposed by a great biologist long time ago (Wright, 1932). Similar ones have been repeatedly and independently proposed in biology since then. It lies at the core in the formulation of evolutionary dynamics, the foundation of biology. It not only corresponds to the energy function in physical science (Ao, 2008), also is a Lyapunov function in control theory of engineering (Ao, 2005a; Haddad and Chellaboina, 2008), two fields strongly associated with systems biology. The idea is also closely related to the mathematical theory of large deviation (Feng and Kurtz, 2006; Varadhan, 2008). Nevertheless, this concept has suffered certain conceptual and theoretical problems, and has not been nearly forgotten by modern molecular biologists. Recently, in studying the stability and robustness of phage lambda genetic switch we accidentally discovered the key to solve those conceptual and theoretical problems, and have shown its usefulness in systems biology.

In the present overview I will give a short presentation on this important recent progress. I will start with discussions on five known ideas on the adaptive landscape in Section II. Some of them have been very successful, while others are still in the metaphoric stage. Then I discuss the central issue on the adaptive landscape in Section III: its existence and consistency in biological sciences and beyond. In section IV I will discuss associated open problems in general terms, and conclude in Section V. I hope I will be able to convey the importance and powerfulness of the adaptive landscape in network modeling, and the open problems would draw the attention of quantitative biologists.

## II. Biological roots of adaptive landscape

In this section we show the usefulness, metaphorically and/or quantitatively, of the adaptive landscape concept in various subfields in biology, manifested by its numerous independent originations. It captures an essential part of dynamics in biological processes.

### II.1. Population genetics.

The concept of adaptive landscape appears to be basic to describe evolutionary dynamics in biology. It has been repeatedly and independently come up in various biological contexts. The most famous proposal, arguably the first, may be due to S. Wright, proposed in 1932 (Wright, 1932, 1988). There are several ways to express the key idea from today's perspective. Wright already used two opposite representations in his original proposal, in my opinion: the probabilistic distribution for an ensemble--a collection of large agents, and the trajectory for a single agent (Fig. 1). Such multiple representations have caused a sustained confusion in the literature. In addition, Wright apparently recognized the generality of his proposal and had applied his concept far beyond the initial intended population genetics, a move seemed not been widely appreciated. We will discuss related issues in next two sections, particularly that of a "flat" landscape situation. Nevertheless, such generality is vindicated by similar proposals from fields very far from population genetics to be discussed in rest of this section, and by its influence in other fields. For example, his concept of fitness landscape has been routinely used in statistical physics known as fitness landscape, and many researchers there may not know its originator. Furthermore, conceptual problems encountered by Wright's adaptive landscape are also shared in different forms in other situations.

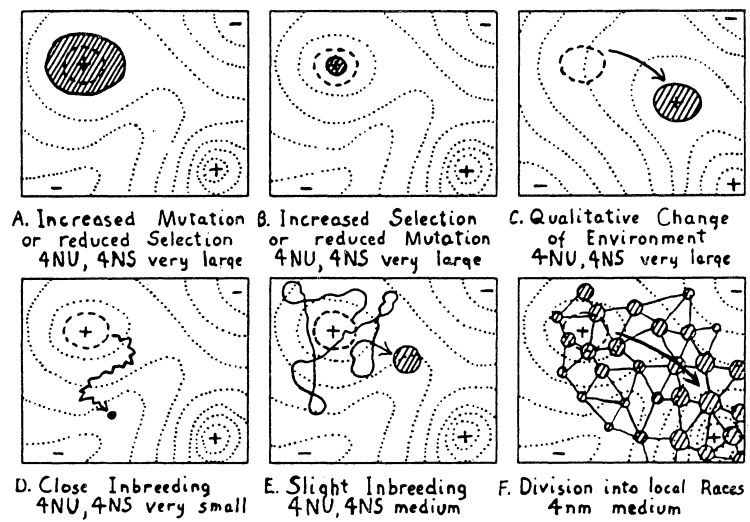

**Figure 1**. Wright's adaptive landscape schematically represented with a portion of the multidimensional landscape of genotypes of a single population with potential (fitness) contour, and his shifting balance theory. The most likely initial probability distribution of individual agents of a large population is indicated by the shaded area (Fig. 1A). There is a first increase in average fitness, represented by the shrinking shaded area (Fig. 1B). After some waiting time, the population transverses across a saddle configuration to another likely better fitness peak (Fig. 1C). Fig. 1C-D are the trajectory view of the same process to emphasize the small population limit, and Fig. 1F indicates the discreteness nature of space. (From S. Wright, 1932, Fig. 4)

### II.2. Developmental biology.

Independent of S. Wright, a similar concept was proposed by CH Waddington in 1940 and elaborated further in 1957 explicitly as a metaphor, at least so far, to understand the developmental process, known as developmental landscape (Waddington, 1957) (Fig. 2). The present author has not been aware of any realization of such developmental landscape directly corresponding to dynamical data. The interaction between developmental processes to both environmental and genetic factors can be discussed in Waddington's landscape (Waddington, 1957; Slack, 2002). The profound symmetry breaking idea embedded in such landscape captures an important aspect of developmental processes. It also vividly and graphically depicts the robustness and plasticity, the modern interpretations of Waddington's canonization.

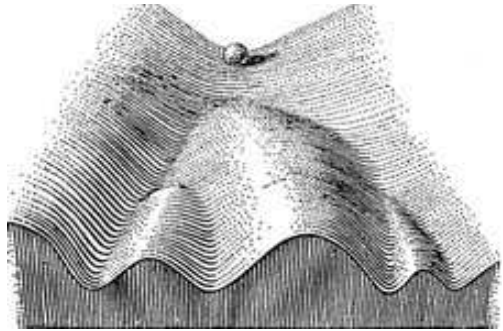

**Figure 2**. Waddington's developmental landscape. The ball represents a cell, and the bifurcating system of valleys represents the bundles of trajectories in the functional space. Similar to Wright's shifting balance theory, such landscape indicates the possibility of transitions between different functional states. (From CH Washington, 1957. Fig.4)

## II.3. Gene regulation and genetic switch.

In discussion of zero-none genotypic phenomena M. Delbruck proposed in 1949 (Delbruck, 1949) the possibility of bi-stable states in genetic systems (Fig. 3). He argued that such situations could be formed based on known physics and chemistry principles. A few years later such phenomena were clearly observed experimentally (Novick and Weiner, 1957). Subsequently an innovative and concrete molecular biological mechanism was proposed and further tested experimentally (Jacob and Monod, 1961). Nowadays there exists an extensive quantitative and predictive study of such genetic switch systems (Zhu *et al.*, 2004, 2007; Ptashne, 2004). The bio-switches have been one of building elements in the study of systems biology. This is an example that once the dynamics is explicitly known, not only the adaptive landscape can be constructed quantitatively, various related questions, such as stability, robustness, etc, can be studied and compared with further experimental data.

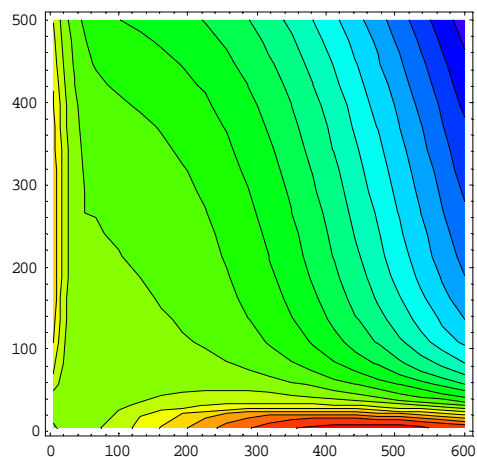

**Figure 3**. Bistable genetic switch. The genetic switch idea was first proposed by M. Delbruck. Perhaps due the brevity of his comment, or, the obviousness (to him) of such idea, M. Delbruck didn't draw a bi-stable landscape. The landscape here is instead for phage lambda genetic switch, taking from Zhu *et al.* (2004). The regimes with dense equal potential lines are two attractive basins, the viable two stable states. The coordinates are the numbers of two proteins

controlling the switch. It appears to be the first quantitative construction of such landscape for genetic switch based on physics, chemistry, and biology principles. (From Zhu *et al*., 2004, Fig. 4b )

## II.4. Neural dynamics and computing.
An analog computer was proposed for the neural dynamics in 1982 by JJ Hopfield. It has inspired a tremendous amount of research activity, reviewed in perspective by Hopfield himself in 1999 (Hopfield, 1999). In this research program the "energy" landscape can actually be constructed quantitatively for many interesting situations. A schematic illustration is given in Fig. 4. The stable states in the adaptive landscape represent the possible solutions from the neural network "computing".

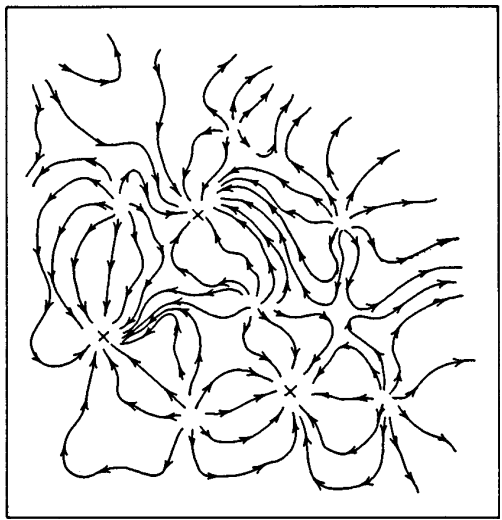

**Figure 4**. Hopfield's landscape for neural dynamics inspired computation in terms of the flow field. The stable points of the flow, marked by x's, are possible answers. To initiate the computation, the initial location in state space must be given. A complex analog computer would have such a flow field in a very large number of dimensions. (From Hopfield, 1999, Fig. 1).

## II.5. Protein folding.
Perhaps the best known recent example of landscape concept is from protein folding research, where such a concept has been playing both the metaphoric and quantitative roles during past two decades. There is no doubt by researchers that such landscape exists. Nevertheless, computing it directly from first principles in physics to detail the folding dynamics of amino acid chains has not been possible, not even in the foreseeable future with faster computers. A coarse-grain averaged landscape capturing the major features of folding dynamics is needed: both as a guidance to test hypotheses and as an intermediate quantitative realization. Indeed, such a landscape concept was proposed 10 years ago (Bryngelson *et al*., 1995; Dill *et al*., 1995), which helps the researchers enormously to such extend the protein folding dynamics has been announced to be solved by an optimistic group of experts (Service, 2008). It may be interesting to note that the various stages embedded in the funnel landscape corroborate well with the modular phenomena widely observed in biological systems.

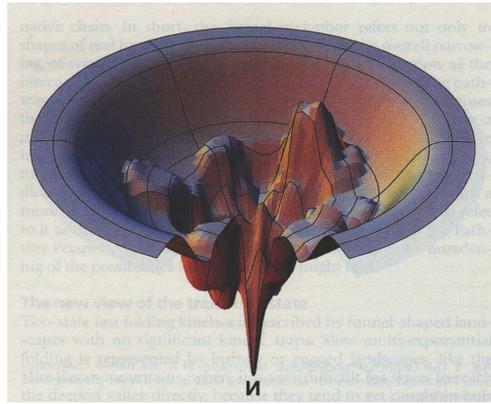

**Figure 5**. Protein folding funnel landscape. It is a down hill run. A folding protein can follow many paths to its most energetically stable native (N) conformation (From Dill and Chan, 1997, Fig.4.). More than ten orders of scale in energy have been involved, showing such idea can be used to link various hierarchical levels. The dynamical flow is somehow opposite to that in Waddington's developmental landscape.

**II.6. Sampling of recent studies.**
An exhaustive list of all studies on dynamics in biology with adaptive landscapes is not possible in this short paper, even though this concept has been somehow "forgot" by modern molecular biologists. Instead, a few recent works in various subfields in biology are given here to indicate the usefulness of the concept.

The construction of adaptive landscape for evolutionary dynamics in the wild has a few successful examples (Endler, 1988; Kauffman, 1993), where ecological factors are obviously important. The evolutionary history of a protein has been illustrated by its motion in a landscape constructible from experimental data (Lunzer *et al*., 2005). The possible evolutionary route for another protein was demonstrated in a similar fashion (Poelwijk *et al*., 2007). In additional to make use of landscape concept in genetic switch study (Morelli *et al*., 2008; Toulouse *et al*., 2005; Yuan *et al*., 2008; Zhu *et al.*, 2004; 2008), similar investigations have been carried out into other important molecular biological processes, such as cell cycle (Wang *et al*., 2006; Zhang *et al*., 2006), signal transduction pathways (Lapidus *et al*., 2008) and cancer genesis and progression (Andrews 2002; Ao *et al*., 2008): The adaptive landscape immediately quantifies important concepts such as robustness and stability, and can be easily understood graphically. The quantitative realization of the concept has been further explored in discrete space (Cao and Liang, 2008; Walczak *et al*., 2005). The roles of such concept have been quantitatively explored in various recent studies (Arnold et al., 2001; Hendry *et al.*, 2006; Holmstrom and Jensen, 2004; Waxman and Gavrilets, 2005). Hence, it is evident that the adaptive landscape concept has been very useful both metaphorically and quantitatively in biology.

**III. Major theoretical progress**

Given such extensive support for the adaptive landscape concept and associated potential function, there is still a strong opposition on its usage ranging from biologists, chemistry, and physics (for a selective survey, see, for example, Ao, 2005; 2008), and not necessarily confined to any particularly subfield in biology. Similar worries can be found in other fields (Ao, 2004). One would naturally wonder the reasons behind those numerous objections and rejections, which cannot be all trivial. Indeed, there were, and still are, a serious set of open problems. In this section problems associated with mathematical and conceptual side, such as its existence in a general setting in the domain of theoretical biology as re-defined recently (Brenner, 1999), are discussed.

**III.1. Origin of the problem and the current solution.**
It was already recognized shortly after Wright's proposal of adaptive landscape that there is a problem associated with a simple of view of evolutionary dynamics as fitness increasing processes: always increasing mean fitness in the adaptive landscape. There are biologically meaningful and realizable cyclic dynamical processes in which the system can repeatedly visit its starting state indefinitely. The challenge is to construct an adaptive landscape for such a process in a consistent manner. It was concluded that the fitness could not exist because when system comes back to its starting state its fitness should not increase. Because of lacking of generic constructive method, this theoretical and mathematical problem remained open till 2004, to the present author's knowledge. This challenge was formulated in most explicitly for the case of limit cycle dynamics: Finding such adaptive landscape with both local and global meanings. This challenge was taken up by the present author. It was found that a proper understanding of adaptive landscape has to make use of the concept of probability or stochasticity, such as embedded in the so-called fundamental theorem of natural selection (Fisher, 1930; Li, 1955). After clarifying the controversies surround the fundamental theorem of natural selection and adaptive landscape, a consistent mathematical formulation for evolutionary dynamics was found (Ao, 2005), in which the quantified adaptive landscape as adaptive potential function is an integral part of the formulation. Specifically, the adaptive landscape for a limit cycle was explicitly and exactly constructed (Ao, 2005; Wang *et al.*, 2008; Zhu *et al.*, 2006). It was found that right at the limit cycle, the adaptive potential function takes same value in the deterministic limit, a "flat" section of adaptive landscape. The dynamics is still possible at the limit cycle because of the existence of a conservative dynamics, a part of dynamics largely overlooked in previous evolutionary studies.

**III.2. Three independent dynamical elements of bionetworks.**
**The general structure**. While it is not intended to give a full account of construction, it would be helpful to show in more quantitative manner what would the related quantities look like. In a typical trajectory description, the dynamical equation would take the form of standard stochastic differential equation. With an appropriate time scale, it would read,

$$\partial_t \mathbf{q} = \mathbf{f}(\mathbf{q}) + N_I(\mathbf{q}) \zeta(t) \quad , \quad (1)$$

where **f** and **q** are n-dimensional vectors and **f** a nonlinear function of **q** and possible time t, too. This implies that the network has n-node. The network variable of i-th allele is represented by $q_i$. Depending on the situation under consideration, the quantity **q** could, alternatively, be the populations of n species in ecology, the numbers of n proteins, or, the n coordinates in physical sciences. All quantities in this paper are dimensionless: They are assumed to be measured in their own proper units. The collection of all q forms a real n-dimensional phase space. The noise $\zeta(t)$ is explicitly separated from the state variable to emphasize its independence, with l components. It is a standard Gaussian white noise function with zero average, and the covariance matrix element $<\zeta_i(t) \zeta_j(t')> = 2 \delta_{ij} (t-t')$, and i,j = 1,2, …, l . Here $<>$ denotes the average over the noise variable $\{\zeta(t)\}$, to be distinguished from the average over the distribution in phase space below. The variation is described by the noise term in Eq.(1) and the elimination and selection effect is represented by the force **f** . A further description of the noise term in Eq.(1) is through the diffusion matrix $D(\mathbf{q})$, which is defined by the following matrix equation $N_I^\tau(\mathbf{q}) N_I(q) = D(\mathbf{q})$ with $N_I$ an n x l matrix and $N_I^\tau$ its the transpose, which describes how the system is coupled to the noisy source. This is a generalization of fundamental theorem of natural selection (Fisher, 1930) in population genetics. By construction the diffusion matrix D is both symmetric and nonnegative. For the dynamics of the state vector **q**, all that is needed from the noise term in Eq.(1). Nevertheless, it had been not clear how a consistent potential function, that is, a consistent adaptive landscape, could be constructed out of Eq.(1).

Our recent work has shown that there is indeed a consistent way to construct the potential function from Eq.(1) (Ao, 2004; Kwon *et al.,* 2005; Yin and Ao, 2006). We refer the readers to original papers for details. Here we simply quote the results. Eq.(1) can be transform into following equation,

$$[ A(\mathbf{q}) + T(\mathbf{q}) ] \partial_t \mathbf{q} = \partial_\mathbf{q} \psi(\mathbf{q}) + N_{II}(\mathbf{q}) \zeta(t) , \qquad (2)$$

where the noise is from the same source as that in Eq.(1). The parameter denotes the influence of non-dynamical and external quantities. It should be pointed out that the potential function $\psi(\mathbf{q})$ may also implicitly depend on time t, if the original **f** does. The friction matrix $A(\mathbf{q})$ is defined through the following matrix equation $N_{II}^\tau(\mathbf{q}) N_{II}(\mathbf{q}) = A(\mathbf{q})$. The anti-symmetric matrix $T(\mathbf{q})$ representing the dynamics which would not change the value of $\psi(\mathbf{q})$. Those matrices are related to the diffusion matrix via $[ A + T ] [D + Q ] = 1$, and

$$[ A + T ] D [ A - T ] = A , \qquad (3)$$

with Q another anti-symmetric matrix. The connection to the deterministic force **f** (**q**) is via an anti-symmetric matrix equation

$$\partial_\mathbf{q} \times \{ [ A(\mathbf{q}) + T(\mathbf{q}) ] \mathbf{f}(\mathbf{q}) \} = 0 , \qquad (4)$$

with $\partial_\mathbf{q} \times$ the wedge differentiation in higher dimensions and

$$\partial_\mathbf{q} \psi(\mathbf{q}) = [ A(\mathbf{q}) + T(\mathbf{q}) ] \mathbf{f}(\mathbf{q}) . \qquad (5)$$

Thus, Eq.(4) is precisely the potential function condition. It can be found via the integration over Eq.(5), independent of the integration routes connecting initial and final points. All A,T,Q, as well as the potential function $\psi(\mathbf{q})$ are uniquely determined by the diffusion matrix and the deterministic force **f** (**q**). It has been shown that the potential function $\psi(\mathbf{q})$ indeed play precisely the role of adaptive landscape, as it has the role of "energy" function in physical sciences, thus quantifies the adaptive landscape, such as those in Fig.1-5.

Because of the constraint between the noise and the diffusion matrix, there are only three independent dynamical quantities in Eq.(2): the potential function, the friction or diffusion matrix, and the anti-symmetric matrix represented by T. In terms of probability distribution function $\rho(\mathbf{q},t)$ in the n-dimensional state space, the dynamics equation will take the form of the Fokker-Planck equation, or, in the discrete space situation, the Master equation. In its abstract form, such equation can be can be written as

$$\partial_t \rho(\mathbf{q}, t) = \mathbb{L} (D, Q, \psi(\mathbf{q}) ) \rho(\mathbf{q}, t) . \qquad (6)$$

A technical discussion of such three independent dynamical elements can be found in Ao (2008).

**An example.** In one dimension, above construction is relatively easy, because the anti-symmetric matrix does not exit. Given the deterministic force **f** (**q**) and diffusion constant (not a matrix function), the potential function can be analytically obtained, already known to physicist Langevin 100 years ago, and its additional mathematical subtle points have been fully recognized since 1950's. A related discussion can be found in standard references on stochastic differential equations [see, for example, references cited in Ao *et al.*, 2007]. The real important situation is in higher dimensions, where there is no time reversibility, or, the detailed balance condition is balance. Let's us illustrate it with a generic two dimensional case, which would include the realistic situations such as phage lambda genetic switch. Following presentation closely follows that in Ao, 2005b.

By introducing an auxiliary matrix $G = D + Q = 1 / [A + T]$, the equations corresponding to Eq.(3) and (4) are,

$$G + G^{\tau} = 2 D, \qquad (7)$$

and $\partial_q \times \{ G^{-1} \mathbf{f}(\mathbf{q}) \} = 0$. In many realistic situations, such as in the phage lambda genetic switch, functions involved are smooth, which allows the possibility to use the gradient expansion, which turns the differential equations of Eq.(5) into an algebraic equation. For the first order, in terms of Jabocian of deterministic force $\mathbf{f}(\mathbf{q})$, which defines as, $F_{11}(\mathbf{q}) = \partial_1 f_1(\mathbf{q})$, $F_{12}(\mathbf{q}) = \partial_2 f_1(\mathbf{q})$, $F_{21}(\mathbf{q}) = \partial_1 f_2(\mathbf{q})$, $F_{22}(\mathbf{q}) = \partial_2 f_2(\mathbf{q})$, the potential condition becomes very simple,

$$G F^{\tau} - F G^{\tau} = 0. \qquad (8)$$

Thus, the auxiliary matrix G has been readily solved analytically and explicitly in this two dimensional case under the gradient expansion: three linear equations in Eq.(7) and one linear equation in Eq.(8). Higher order gradient expansion is a successive procedure of solving such linear algebraic equations to the desired order. Once G is known, the potential function is $\psi(\mathbf{q}) = \int d\mathbf{q} \cdot [ G^{-1} \mathbf{f}(\mathbf{q}) ]$.

In the present author's view, a consistent theoretical formulation adaptive landscape has been now here since 2005 (Ao, 2005a): Major theoretical and mathematical associated with such stochastic evolutionary dynamics are solved, with respect to the construction (Ao, 2004; Kwon *et al.*, 2005), the connection to other stochastic methods (Yin and Ao, 2006; Ao *et al.*, 2007), and to relation to thermodynamics (Ao, 2008), along with some practical computing techniques (Ao, 2004; Kwon *et al.*, 2005). Nevertheless, there are still theoretical issues merited further discussions.

### III.3. Adaptive landscape beyond its metaphoric role
The most clearly formulated "counterexample" in the literature against adaptive landscape was on limit cycle (Ao, 2005a). Other voiced objections, known to the present author, are all verbal and vague. If some would be valid, it would be helpful to formulate them explicitly and quantitatively. If the present author's assessment of the situation correct, based on recent theoretical constructions (Ao, 2005; Kwon *et al.*, 2005; Yin and Ao, 2006) and applications to various concrete biological problems as discussed above, the answer to above question is indeed positive on the theoretical level.

### III.4. What is the effect of noise or drift?
It has been noticed that there are some situations that the landscape seems ill-defined (Poelwijk *et al.*, 2006; Ao, 2005a). From the general theoretical framework (Ao, 2004; Kwon *et al.*, 2005; Yin and Ao, 2006) a sure way to generate such uncertainty is in the no-noise limit. Mathematically, this limit is singular, that is, a procedure for this limit has to be specified. For a complete deterministic dynamics, a potential function not only exists, there can be infinite many of them, perhaps more than what one would like to have. The noise or drift, including both genetic and environmental, is shown to be able to affect the adaptive landscape in a profound and quantitative manner: It is the quantity needed to make the adaptive landscape unique in a given situation, and can determine the outcome of dynamics. As discussed recently (Ao, 2004; Kwon *et al.*, 2005; Yin and Ao, 2006), in the continuous trajectory representation of the dynamics, evolutionary dynamics can be uniquely decomposed into four components, with adaptive landscape, noise, adaptive (non-conserving) and conservative dynamics. The noise is connected to the adaptive dynamics by the F-theorem, an idea already embedded in Fisher's fundamental theorem of natural selection (Fisher, 1930; Li, 1955). Thus, there are actually three independent dynamical elements for a general evolutionary process. In the probability distribution representation of the dynamics such as in the form of Fokker-Planck equation (Ao, 2004; Kwon *et al.*, 2005; Yin and Ao, 2006) or discrete Master equation (Ao, 2008), three independent dynamical elements indeed completely specifies the dynamical equation. In particularly, a complete evolutionary dynamics with "flat" adaptive landscape is possible. This way of decomposing dynamics has also been discussed in detail qualitatively in literature, though the conservative dynamics has not been explicitly expressed (Ao, 2005a). There is an interesting analogy here. Let's recall the situation of the fundamental theorem of arithmetic in high school mathematics: every natural number greater than one can be uniquely written as a product of prime numbers. Parallel to it, this unique decomposition in evolutionary dynamics may be named the fundamental theorem of dynamics.

Evidently the noise, variation, or drift is indispensible in bionetwork dynamics. An important difference between the fundamental theorem of natural selection (Fisher, 1930) and its recent extension as the F-theorem (Ao, 2005a) should be pointed out here. In the formulation of F-theorem no reference to adaptive potential function (or fitness landscape) is used. This suggests that the F-theorem is valid for any shape of adaptive landscape, near or far away for a local peak. In particular, this would imply that it applies to the flat landscape, too. When applying it to a stationary state near a peak, which would be reached after a perturbation for sufficient long time, there should be no change in average value of potential function, while a finite variation would exist. This appears to be what has been observed experimentally, consistent with the F-theorem in the light of present discussion, but has been interpreted by many biologists as evidence against the fundamental theorem of natural selection.

To summarize the situation in theoretical biology, there exists now a consistent and quantitative formulation of dynamics of any bionetwork, so long as its mathematical description is in the form of usual stochastic processes. The adaptive landscape, quantified as a potential function, has been naturally incorporated, which gives a graphical view on the global dynamics.

**IV. Open questions in broad perspective**

There are other unclear and difficult issues associated with the concept of adaptive landscape. In this section three levels of problems are used to classify them, and the third one is of major concerns of working quantitative biologists.

**IV.1. Conceptual domain.**
The first type is at philosophical and linguistic level. It is known that there exist numerous definitions of fitness and associated adaptive landscape. Even if some of them may be equivalent to each other and may be correct, it cannot be so for all of them. This problem has already been noticed long ago (see for example, Ao, 2005), and its proper discussion is beyond the scope of present paper, though a critical evaluation of those in literature is clearly needed. Nevertheless, one cannot dispel the feeling that many discussions on this issue in the literature, by biologists, philosophers and historians, indicate most traditional biologists have not walked out of the shadows of past giants (Haldane, 2008).

The existence of adaptive landscape implies a profound concept in nature: an order, or a priority, in any dynamical process. This quantity is obviously connected to the Hamiltonian, or energy function, in physical sciences (Ao, 2005a; 2008). In addition, discussion in previous section demonstrated an intrinsic connection between the adaptive landscape and the noise or drift, which appears to be most eminent and distinct in the evolutionary dynamics, exemplified by S. Wright and R.A. Fisher. Further explorations on the implications of those important relationships are desirable at the level beyond biology, to lay a firm theoretical foundation for systems biology in particular and biology in general.

**IV.2. Theoretical domain.**
The second type of problems is in the theoretical biology domain: whether or not the adaptive landscape necessarily exists in the general framework of theoretical biology. Various related issues were discussed in Section III. When the dynamics is known quantitatively, generically three independent dynamical components would be uniquely found, built upon the insights of Wright and Fisher. The adaptive landscape quantified as the potential function is one (Ao, 2005a; 2008). If this problem had not been addressed properly, the debate on adaptive landscape would easily persist indefinitely, and its usage in real problems will be greatly reduced. One cannot dispel the feeling that this problem has not been considered seriously enough by theoretical biologists, and it is not the job of philosophers and historians unless their hands become wet.

Another important issue is the discrete vs continuous descriptions, exemplified by a current false prevalent argument that the Master equation is more fundamental in biology that the Fokker-Planck equation (Mehta *et al.* 2008). The fact is that the Master equation has usually been derived based on Fokker-Planck equation. It is true for a given biological problem we need appropriate tools: nobody wants to describe a bacterium starting from string strings even if such theories exist. In the domain two different methods both are valid, same prediction should be expected. If not, there must be an inconsistency in theoretical treatment, which needs to be located. The possibility of generating very different results from apparently same equation has been well known (Ao *et al*., 2007). The associated issue consists of another open problem.

The importance of potential function has been known in another important field, the dynamical systems theory for nonlinear systems. There has been a debate on the general existence of potential function, nevertheless, in the form of gradient systems vs vector field flow systems. To the best knowledge of the present author, such debate has not been closed in mathematics [Holmes, 2005]. The methodology reviewed here appears to provide a key insight to close this debate in dynamical systems.

The implications and applications of the novel construction of Lyapunov function in control theory of engineering has not been full explored yet (Ao, 2005a; Milton *et al*., 2008; Haddad and Chellaboina, 2008), another open research program.

**IV.3. Empirical domain.**
The third type is the concern of practicing biologists: how to construct and to use the adaptive landscape. For a given problem, even if all of us agree on the existence of adaptive landscape, its exact construction and analysis can be a daunting task. The protein folding problem discussed in Section II is one of such ideal example (Bryngelson *et al*., 1995; Dill *et al*., 1995; Frauenfelder *et al.*, 1991; Wade, 2005). While the funnel type landscape is quantitatively crude and metaphoric, it provides a global picture on how folding dynamics are unfolding (Frauenfelder *et al*., 2006), and, step by step under its guidance solutions and a better quantitative understanding are reached (Service, 2008; Thomas *et al.,* 2005).

In situations where the construction of adaptive landscape is still unclear, such as the developmental landscape, the metaphoric nature of the concept has been enormously helpful (Lunzer *et al.*, 2005; Slack, 2002; Wright, 1988.). Even in this situation, thanks to many biologists' decades of dedicated work, we are close to have a complete set of quantitative set ready for meaningful and predictive mathematical modeling (Oliveri *et al*., 2008). In the cases such as phage lambda genetic switch, the landscape can be constructed quantitatively, which leads to solution of outstanding stability puzzle, along with quantitative predictions (Zhu *et al*., 2004; 2007). The constructed landscape can also be useful in improving existing stochastic simulation methods (Hemberg and Barahona, 2007). Thus, new techniques for its construction are needed.

The most difficult situations are those where the metaphor is not enough and a quantitative description is immediately needed, but its clear construction is unknown at beginning. The relevant construction has to be found by trial and error in research. This has been the situation encountered in population genetics (Endler, 1985), and is typical in biology, such as in the cancer dynamics (Ao *et al.*, 2008). This may not be surprising. The examples in Section II amply illustrate the hierarchical structure in biology: There many hierarchical layers in the gap between the geno and pheno types (Ellegren and Sheldan, 2008), in addition to ecological and environmental factors. It is seldom the situation that theoretical models work at its first try, particularly those in population genetics striving to span over this huge gap. The present author certainly shares many of the skeptic's concerns on the use of adaptive landscape right at the genetic level. Such models may well be oversimplified, as already discussed by Haldane in a similar situation (Haldane, 2008), but it is not the reason to eliminating them, because it gives the information for improvement. The Wright's adaptive landscape is, however, a unifying quantity applicable to multiple hierarchical layers. Evidently, practicing biologists, the present author included, have to work hard in any foreseeable future.

**V. Conclusion**

Based on above presentation, we may conclude that the adaptive landscape is one of the central concepts in the modeling of bionetwork dynamics. It can, and has been used, to describe bionetwork dynamics across many hierarchical layers. While important progress has been made during in the first decade of 21$^{st}$ century, open problems at all levels still exist. The selected open problems here serve to get the attention of more quantitative biologists. Together we will move the field forward.

**Acknowledgements:** Insightful discussions with D. Galas, L. Hood, and M.E. Lidstrom at various stages of the research program are of great helpful in shaping my outlook. I also thank C. Kwon, H. Qian, D.J. Thouless, L. Yin and X.-M. Zhu for fruitful collaboration which make the present review possible.
This work was supported in part by a grant from USA National Institutes of Health No. K25-HG002894-05 and by 985 Project from Shanghai Jiao Tong University.
(*Note on literature*: It is impossible to give a complete list of all relevant references due to the emerging and rapid development of this research program. Nevertheless, I wish the somewhat subjective selected references here are useful and would lead to larger part of literature. I apologize to those whose important works are not here due to my oversight. I do wish to be corrected when serious omission/negligence would occur (aoping@u.washington.edu)).